\documentclass{article}
\usepackage{spconf,amsmath,graphicx,adjustbox,booktabs,multirow}
\PassOptionsToPackage{hyphens}{url}\usepackage[hidelinks]{hyperref}
\usepackage{xcolor}
\usepackage{listings}
\usepackage[linesnumbered,ruled,vlined]{algorithm2e}

\title{Wespeaker baselines for VoxSRC2023}
\name{Shuai Wang, Chengdong Liang, Xu Xiang, Bing Han, Zhengyang Chen, Hongji Wang, Wen Ding}
\address{
  Wespeaker Team, WeNet Open Source Community \\
  wangshuai@cuhk.edu.cn
  }

\begin{document}

\maketitle
 
\begin{abstract}

This report showcases the results achieved using the wespeaker toolkit for the VoxSRC2023 Challenge. Our aim is to provide participants, especially those with limited experience, with clear and straightforward guidelines to develop their initial systems. Via well-structured recipes and strong results, we hope to offer an accessible and good enough start point for all interested individuals. In this report, we describe the results achieved on the VoxSRC2023 dev set using the pretrained models, you can check the CodaLab evaluation server for the results on the evaluation set. \textbf{Any feedback and contribution are always welcome}

\end{abstract}

\section{The VoxSRC Challenges}

The VoxSRC (VoxCeleb Speaker Recognition Challenge) is an annual competition that focuses on the task of speaker recognition using the VoxCeleb dataset. Speaker recognition is a field within audio processing that aims to identify and authenticate individuals based on their unique vocal characteristics.

The VoxSRC Challenge serves as a platform for researchers and practitioners to showcase their advancements in speaker recognition technology. It provides a standardized evaluation framework, allowing participants to compare their methods and algorithms against each other.

VoxSRC 2023 consists of four tracks, which are consistent with the previous year's competition. Tracks 1, 2, and 3 are dedicated to speaker verification, where participants are required to determine whether two speech samples originate from the same person. The evaluation for Tracks 1 and 2 will be conducted on the same dataset, with Track 1's training data restricted to the VoxCeleb2 dev set, while participants can freely use any data for Track 2. 

Track 3 aims to promote domain adaptation research, providing an evaluation set from another domain (CnCeleb dataset). It includes a large set of unlabelled data and a small set of labelled data from the target domain to serve as the adaptation data. The objective is to address the challenges of adapting speaker verification models to different domains.

On the other hand, Track 4 focuses on speaker diarisation, challenging participants to accurately segment multi-speaker audio into distinct portions that correspond to individual speakers. This track addresses the problem of determining ``who spoke when'' in a given audio recording.

\begin{table*}[!htb]
\centering
\caption{\label{tab:vox_basic} Table 1 presents the results achieved using different architectures on the VoxCeleb dataset and VoxSRC2023 development set. The "dev" portion of part 2 is used as the training set. These results serve as a baseline for Track 1 and Track 2 in the VoxSRC challenge. The $p_{target}$ value is set to $0.05$ for the voxsrc23\_val dataset, while for other datasets, it is set to $0.01$.}
\begin{tabular}{ccccccccc}

\toprule
\multirow{2}{*}{Architecture} & \multicolumn{2}{c}{voxceleb1\_O} & \multicolumn{2}{c}{voxceleb1\_E} & \multicolumn{2}{c}{voxceleb1\_H}  & \multicolumn{2}{c}{voxsrc23\_{val}} \\
& EER(\%) & minDCF & EER(\%) & minDCF & EER(\%) & minDCF & EER(\%) & minDCF \\\midrule
CAM++ & 0.654 & 0.087 & 0.805 & 0.092 & 1.576 & 0.164 & 3.899 & 0.211\\\midrule
ECAPA-TDNN & 0.728 & 0.099 & 0.929 & 0.100 & 1.721 & 0.169 & 4.392 & 0.228\\ \midrule
ResNet34 & 0.723 & 0.069 & 0.867 & 0.097 & 1.532 & 0.146 & 3.660 & 0.213\\ \midrule
ResNet221 & 0.505 & 0.045 & 0.676 & 0.067 & 1.213 & 0.111 & 2.991 & \textbf{0.168}\\ \midrule 
ResNet293 & \textbf{0.447} & \textbf{0.043} & \textbf{0.657} & \textbf{0.066} & \textbf{1.183} & \textbf{0.111} & \textbf{2.867} & 0.169\\ \bottomrule
\end{tabular}
\vspace{-0.1cm}
\end{table*}

\section{WeSpeaker: Speaker Embedding Toolkit for Research \& Production}
\label{sec:wespeaker}
\subsection{Open-source speech processing toolkits}
In the field of speech processing, the research community has made significant contributions to the open-source domain. Initially, toolkits such as HTK (Hidden Markov Model Toolkit)~\cite{young2002htk} and Kaldi~\cite{povey2011kaldi} played a pivotal role in enabling researchers and industry applications. However, the emergence of deep learning toolkits like PyTorch and TensorFlow has brought about a shift in the landscape.

Recently, PyTorch-based toolkits such as SpeechBrain~\cite{ravanelli2021speechbrain} and ESPnet~\cite{watanabe2018espnet} have gained popularity due to their user-friendly interfaces and support for rapid prototyping, making them accessible to new researchers. While these toolkits serve a broad range of applications, Wenet stands out by focusing specifically on end-to-end speech recognition. Its primary aim is to bridge the gap between research advancements and practical deployment in real-world scenarios.

\subsection{Wespeaker}
 In~\cite{wang2023wespeaker}, we introduced Wespeaker, a speaker embedding learning toolkit designed for research and production purposes. Wespeaker is characterized by its lightweight code base and emphasis on high-quality speaker embedding learning, demonstrating impressive performance on multiple datasets. While prioritizing accessibility for researchers, Wespeaker also incorporates deployment codes that are compatible with both CPUs and GPUs, thereby facilitating the integration of research findings into practical production systems.

 \subsection{Design principles}
 As mentioned in previous section, there are different speech toolkits which include speaker embedding learning functions, our proposed wespeaker stands out for its simpliness, effectiveness and deployment friendliness. THe design principles are as follows,
 \begin{itemize}
    \item \textbf{Light-weight}: Wespeaker is designed specifically for deep speaker embedding learning with clean and simple codes\footnote{If you are interested in other tasks such as ASR, KWS, TTS, etc. We have different speficifcly designed toolkits for different tasks, please visit \url{https://github.com/wenet-e2e} for more details}. It is purely built upon PyTorch and its ecosystem, and has no dependencies on Kaldi~\cite{povey2011kaldi}.

    \item \textbf{Production oriented}: All models in Wespeaker can be easily exported by torch Just In Time (JIT) or as the ONNX format, which can be easily adopted in the deployment environment. Sample deployment codes are also provided.
 \end{itemize}
\subsection{Supported functionalities}
Wespeaker supports different popular speaker embedding learning models, margin based softmax training objectives and several pooling functions. 
 \begin{figure}[!htb]
  \centering
  \vspace{0cm}
\includegraphics[width=\linewidth]{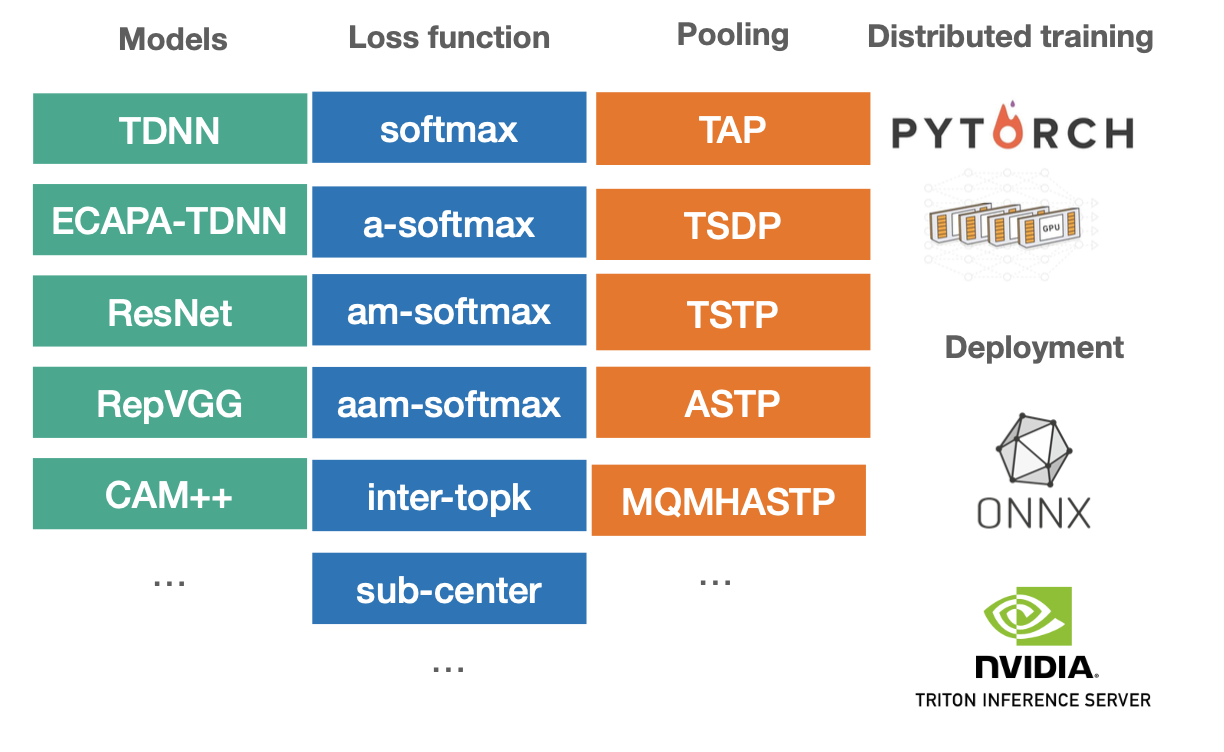}
  \caption{Supported functions in Wespeaker}
  \label{fig:uio}
  \vspace{-0.3cm}
\end{figure}

\textbf{Model Architectures}
\begin{itemize}
    \item \textbf{TDNN} based x-vector~\cite{snyder2018x}, this is a milestone work that leads the following deep speaker embedding era.
    \item \textbf{ResNet} based r-vector and its deeper version, this is the best system of VoxSRC 2019~\cite{zeinali2019but} and CNSRC 2022~\cite{chen2022sjtu}.
    \item \textbf{ECAPA-TDNN}, a modified version of TDNN, this is the champion system of VoxSRC 2020~\cite{desplanques2020ecapa}.
    \item \textbf{RepVGG} decouples the training time and inference time architecture, resulting in good performance and inference speed. This is the best system of VoxSRC 2021 \cite{zhao2021speakin}.
    \item \textbf{CAM++}, a modified a densely connected time delay neural network (D-TDNN) that utilizes a context-aware masking mechanism. It also incorporates a novel multi-granularity pooling technique to capture contextual information at various levels.
\end{itemize}

\textbf{Pooling functions} 

Pooling functions aggregate frame-level features into segment-level representations, where Wespeaker supports the statistics-based and attention-based ones.

\textbf{Loss functions} 

Loss functions play a crucial role in deep speaker embedding learning. We provide support for various types of loss functions, including the standard softmax cross-entropy loss, as well as different margin-based variants~\cite{hajibabaei2018unified,xiang2019margin}. These variants include A-softmax~\cite{liu2017sphereface,huang2018angular}, AM-softmax~\cite{wang2018additive}, and AAM-softmax~\cite{deng2018arcface}.

In addition to supporting different loss functions, we also provide support for commonly used techniques such as the inter-topk and sub-center algorithms. These techniques aim to enhance the discriminative ability of the learned embeddings by considering specific subsets of samples within a mini-batch or using sub-centers to improve intra-class compactness.

\textbf{Scoring back-ends}

In the toolkit, we offer a basic implementation of the two-covariance Probabilistic Linear Discriminant Analysis (PLDA). We encourage users to explore different adaptation methods~\cite{povey2011kaldi, alam2018speaker, lee2019coral+, bousquet2019robustness} with PLDA for the Track 3 cross-domain evaluation condition.
\subsection{Easy hands-on}

We have included pretrained models in the toolkit to assist users in quickly verifying results on relevant datasets. However, we would like to emphasize that \textbf{\color{red}we DO NOT recommend users to solely submit results based on the provided single systems}. We encourage users to explore different methods of combining systems, either among the models we provide or with ones trained by themselves.

We provide the python binding for wespeaker for users to quickly try the pretrained models, further details could be found on the project webpage \url{https://github.com/wenet-e2e/wespeaker/tree/master/runtime/binding/python}

With the wespeakeruntime package installed, you can easily extract embeddings from WAV files specified in the wav.scp file and save them into embed.ark using the following code:

\begin{lstlisting}
import wespeakerruntime as wespeaker
wav_scp_path = "path/to/wav.scp"
embed_ark_path = "embed.ark"

speaker = wespeaker.Speaker(lang='chs')
speaker.extract_embedding_kaldiio(
             wav_scp_path, embed_ark_path
             )
\end{lstlisting}

Moreover, we released several pretrained models as dipicted in Table~\ref{tab:pretrain}, both in pytorch ``.pt'' format and runtime ``.onnx'' format, check \url{https://github.com/wenet-e2e/wespeaker/blob/master/docs/pretrained.md} for details on how to use it.

\begin{table}[!htb]
\caption{\label{tab:pretrain} Pretrained models provided}
\centering
\setlength\tabcolsep{3pt}
\begin{tabular}{c|c|c}
\toprule
Datasets & Languages & Pretrained model          \\ \midrule
VoxCeleb & EN        & CAM++ / CAM++\_LM          \\\midrule
VoxCeleb & EN        & ResNet34 / ResNet34\_LM \\ \midrule
VoxCeleb & EN        & ResNet152\_LM                  \\ \midrule
VoxCeleb & EN        & ResNet221\_LM                 \\ \midrule
VoxCeleb & EN        & ResNet293\_LM               \\ \bottomrule   
\end{tabular}
\end{table}

\section{Results}

\subsection{Track 1 \& 2}

The results on the VoxCeleb1 evaluation dataset and the VoxSRC 2023 development set are presented in Table~\ref{tab:vox_basic}. The models employed for these evaluations are specified in Table~\ref{tab:pretrain}.

\subsection{Track 3}
There are various technology roadmap for unsupervised domain adaptation, and we only provide the results of the voxceleb pre-trained model in Table~\ref{tab:cnceleb_results}.  

\begin{table}[!htb]
\footnotesize
\setlength\tabcolsep{3pt}
\centering
\caption{\label{tab:cnceleb_results} Results on the validation set of Track3. The $p_{target}$ value is set to $0.01$.}
\begin{tabular}{cccc}
\toprule
Architecture & Mean Normalization & EER(\%) & minDCF \\ \midrule
\multirow{2}{*}{ResNet34}  & N &  14.570 & 0.617 \\ \cmidrule{2-4}
  & Y &  11.395 & 0.594 \\ \bottomrule
\end{tabular}
\vspace{-0.2cm}
\end{table}

\subsection{Track 4}
We used the open-source 
pyannote~\cite{Bredin2020}
toolkit as our Voice Activity Detection (VAD) system~\footnote{This is different from the silero VAD used in ~\cite{wang2023wespeaker}}. The ResNet34\_LM model was adopted as the speaker embedding extractor. For speaker clustering, we implemented the spectral clustering algorithm and adapted it specifically for the diarization task. The results on the VoxConverse dev and test sets are shown in Table~\ref{tab:voxconverse_results}\footnote{Dev and test set of VoxConverse are used as the validate set for the VoxSRC23 Track 4}.

\begin{table}[!htb]
\footnotesize
\setlength\tabcolsep{3pt}
\centering
\vspace{-0.3cm}
\caption{\label{tab:voxconverse_results} Results on the VoxConverse dev and test sets}
\begin{tabular}{ccccc}
\toprule
Test set & MISS(\%) & FA(\%) & SC(\%) & DER(\%)  \\ \midrule
VoxConverse dev  & 2.7 & 0.2 & 1.8 & 4.8  \\ \midrule
VoxConverse test & 3.2 & 0.7 & 3.0 & 7.0 \\ \bottomrule
\end{tabular}
\vspace{-0.2cm}
\end{table}

\section{Suggestions for performance improvement}
Our primary objective is to offer a robust initial model that serves as a strong starting point for further improvement. We aim to provide researchers with a solid foundation from which they can develop and enhance new algorithms. By supplying a sufficiently good initial model, we aspire to facilitate the development of novel methodologies within the research community. We didn't specifically do optimizations for Track 2, 3, and 4. Instead, we would like to provide some suggestions to provide several potential directions  to work on:

\subsection{Track 2}
\begin{itemize}
    \item Increase Data Volume: Expand the training dataset by adding more data. 
    \item Explore Large Pretrained Models~\cite{chen2022large}: Consider utilizing large pretrained models like WavLM~\cite{chen2022wavlm} to leverage their extensive knowledge learned from vast amounts of audio data. 
    \item Pretrained ASR Model Initialization: Phoneme information has been proven to be beneficial for building speaker verification systems~\cite{wang2019usage}. Consider initializing your speaker embedding model with pretrained Automatic Speech Recognition (ASR) models. Several papers presented during ICASSP 2023 verified the effectiveness~\cite{liao2023towards,cai2023pretraining}.
    \item Hard Mining Strategy: Find confused speakers and add an extra inter-topK penalty on them is an effective way to improve performance in challenges~\cite{chen2022build,zhao2022multi,han2023exploring}. Some of them have already been supported in Wespeaker\footnote{https://github.com/wenet-e2e/wespeaker/pull/115}.
\end{itemize}

\subsection{Track 3}
\begin{itemize}
    \item Distribution Alignment: Employ adversarial training or other strategies to align the distributions of the source and target domains.
    \item Pseudo Label Learning: Utilize clustering algorithms or other methods to assign pseudo labels to unlabeled data from the target domain. It is important to note that these pseudo labels may contain noise, and exploring techniques~\cite{han2023exploring,tao2022self,Han2022SelfSupervisedSV} for training robust systems with noisy labels is a crucial topic.
    \item Unsupervised PLDA Adaptation: Building upon the implemented PLDA codes, you can incorporate PLDA adaptation mechanisms, such as the Kaldi version~\cite{povey2011kaldi} and the CORAL series~\cite{alam2018speaker, lee2019coral+}, to further enhance performance\footnote{A good reference on PLDA adaptation can be found as ~\cite{wang2023generalized}}.
\end{itemize}

\subsection{Track 4}
\begin{itemize}
    \item VAD Tuning. Currently, the errors caused by VAD is still quite high, improving the VAD might be a good choice.
    \item We only include the basic clustering algorithms here, you can try more alogrithms and reclustering methods such as VBx~\cite{diez2019bayesian, landini2022bayesian}.
\end{itemize}

\subsection{Results on the evaluation set}
 We submit the best single system (ResNet293) to the CodaLab evaluation server, check the following links for the numbers and rankings.

  \begin{itemize}
     \item Track 1 \url{https://zeus.robots.ox.ac.uk/competitions/competitions/17#results}
     \item Track 2: \url{https://zeus.robots.ox.ac.uk/competitions/competitions/16#results}
     \item Track 3: \url{https://zeus.robots.ox.ac.uk/competitions/competitions/14#results}
     \item Track 4: \url{https://zeus.robots.ox.ac.uk/competitions/competitions/18#results}
 \end{itemize}
 
\section{The story}
The voxceleb dataset is the largest opensource and high-quality dataset for speaker recognition and the wespeaker team has a long history of supporting the voxceleb dataset and voxsrc challenges, the core members have achieved top rankings in previous VoxSRC competitions~\cite{zeinali2019but, xiang2020xx205, chen2022sjtu2}\footnote{VoxSRC2019: 1st place, VoxSRC2020: 2nd place, VoxSRC2022: 3rd place}.

We have observed that there is often a disparity between the results reported in current research papers and the performance achieved in system reports for challenges, even when the training and evaluation data are the same. In order to  provide a reliable starting point for researchers, we initiated wespeaker, aimed at delivering a reliable baseline system and a user-friendly toolkit. Moreover, contributers from the Wenet opensource community helped with the efficient data management techniques that enable scaling to industrial-sized datasets, as well as deployment codes for rapid prototyping in production environments.

Knowing that this would be the final VoxSRC challenge, the Wespeaker team is eager to contribute and support the event by providing an easy-to-use toolkit and baseline systems. We hope more participants can enjoy the challenge and focus on the algorithm improvement, without struggling with basic experimental setups.

\section{Acknowledgement}
We would like to extend our sincere appreciation to the VoxSRC challenge organizers for their invaluable contribution in open-sourcing this remarkable dataset and organizing such meaningful challenges. We would also like to express our gratitude to the wenet open-source community, whose dedication and collective efforts have played a pivotal role in the success and growth of wespeaker. Enjoy the Challenge and Welcome to contribute.

\bibliographystyle{IEEEtran}
\bibliography{mybib}

\end{document}